\documentclass[amsmath,amssymb,showpacs,twocolumn,prl] {revtex4-2} 

\usepackage{color}
\usepackage{graphicx}
\usepackage{amssymb}
\usepackage{amsmath}
\usepackage{mathrsfs}
\usepackage{subfigure}
\usepackage{hyperref}
\usepackage{breakurl}

\begin{document}
\title{Nonlinear propagation and localisation in photonic crystal waveguides}
\author{S. Malaguti}
\begin{abstract}
Anderson predicted that electrons diffused by a disordered potential in doped semiconductors see a metal-to-insulator transition of the material when the disorder is sufficiently high. In this manuscript we demonstrate an equivalent metal-to-insulator transition for nonlinear waves, namely Gap-Solitons, in disordered photonic crystal waveguides. Light localization is described by introducing a new metric able to track the wavepacket center of mass. By statistical averaging this quantity over many realizations of the disorder, we demonstrate that in linear regime the ensemble averaged barycenter matches the localization length $l_{loc}$ as defined in literature. Then, by applying our barycenter method, we prove that for Gap-Solitons the transition from localized to ballistic transport goes faster than the $v_g^2$ law ruling the linear regime. By overcoming this scaling law, improved robustness to disorder of nonlinear waves is demonstrated. 
\end{abstract}

\maketitle

{\it Introduction --}
Recent experiments have demonstrated soliton-like response by lunching high intensity pulses through a photonic crystal waveguide (PhC-WG) \cite{Colman10}. Unlike soliton arising from 1D periodic structures, 2D planar PhC-WG allow to obtain pulse compression and self-transparency over short distances (a few mm) with relative low input power. However, disorder caused by built-in technological imperfections, breaking the lattice symmetry, becomes the cause of detrimental coherent backscattering and related localization phenomena.
Anderson localization [3] is a well-known mechanism attributed to multiple scattering of electrons by a random potential, arising from the wave nature of electrons. As a phenomenon due to wave interference, localization occurs not only for electrons, but also for microwaves, acoustic waves and even Bose-Einstein condensated matter waves.

The aim of this manuscript is to provide a theoretical description of the underlying nonlinear physics in disordered PhC, and demonstrate robustness of solitons against disorder.\

{\bf NLCME system and metrics for disordered PhC-WG}\\
We consider a $L=50\mu$m long line-defect waveguide (common W1) created by removing a row of holes in the $\Gamma K$ direction of a triangular lattice PhC membrane (surrounded by air) with the following realistic parameters: lattice constant $a=480$ nm, holes radius $r=0.30a$, membrane thickness $240$ nm, and bulk refractive index $n=3.17$ (GaInP).
The GaInP bulk material exhibits dominant Kerr effect allowing soliton dynamics, while losses induced by parasitic (linear an nonlinear) effects can safely ruled out.
Relying on the fact that the dispersive properties of the PhC-WG can be properly described trough the anticrossing of the two counterpropagating index- and gap-guided Bloch modes, we introduce the following coupled mode system for the slowly varying envelops $E_{\pm}(Z,T)$ \cite{Malaguti}: 
\begin{equation}\label{eq1}
\begin{split}
&\left(i\frac{1}{V_i}\partial_T + i\partial_Z \right)E_+ +\Gamma E_- +(\gamma_x |E_-|^2 +\gamma_+|E_+|^2)E_+ =0\\
&\left(i\frac{1}{V_g}\partial_T - i\partial_Z \right)E_- +\Gamma E_+ +(\gamma_x |E_+|^2 +\gamma_-|E_-|^2)E_- =0
\end{split}
\end{equation}
where the linear coupling coefficient $\Gamma$ accounts for the coupling between the fast $E_{+}$ (index-guided group velocity $V_i \approx c/n$) and the slow $E_-$ (gap-guided velocity $Vg \ll Vi$) modes, being $c$ the vacuum light velocity and $n$ the refractive index of the bulk material.
The linear dispersion evaluated through linear waves $\exp(iKZ-i\omega T)$ in Eq. (\ref{eq1}) has the two branches
\begin{equation}\label{eq2}
\omega_{\pm} = \frac{K}{2}(V_g-V_i)\pm \frac{1}{2}\sqrt{K^2(V_i+V_g)^2+4V_iV_g \Gamma^2},
\end{equation}
which entail a forbidden indirect gap $|\omega|<\omega_b$, with band-edges $\pm{\omega}_{b}$, and $\omega_b \equiv 2 \Gamma/(V_i^{-1}+V_g^{-1})$ (note that $K$ and $\omega$ represent shifts from the gap center.).
By choosing as best-fit parameters $V_i = 0.8c/n = 7.6\times 10^7$ m/s, $V_g = 0.05c/n = 4.7 \times 10^6$ m/s, and $\Gamma = 0.2n/a= 1.32\times 10^6$ m$^{-1}$ Eq. (\ref{eq1}) correctly captures the main dispersive features of PhC-WG. Assuming a typical value of Kerr coefficients in semiconductors such as GaAlAs, Silicon, or GaInP, we set the nonlinear coefficients $\gamma_+ = \gamma_- = \gamma_x/2 \equiv \gamma$, with $\gamma = 920$ (Wm)$^{-1}$ obtained for a nonlinear index $n_{2I} = 6\times 10^{-18}$ m$^2$/W \cite{Colman10}.
We address the disorder by introducing both, linear and nonlinear random coefficients: $\Gamma(z) = \bar{\Gamma} + \Psi(z)$, $\gamma(z)= \bar{\gamma}+\Psi_{\gamma}(z)$, with mean values $\bar{\Gamma} \equiv \Gamma$, $\bar{\gamma} \equiv \gamma$, and where $\Psi(z)$ and $\Psi_{\gamma}(z)$ are, for each fixed $z$, random Gaussian process with zero mean value and variance $<\Psi^2>= \sigma_{\Psi}^2$, $<\Psi_{\gamma}^2> = \sigma_{\gamma}^2$, respectively. 
Unlike disordered structures investigated so far, as fiber Bragg gratings or evanescently coupled waveguides\cite{Tsoy,Kartashov_OE_2007}, since in PhCs the nonlinearity undergoes a random distributions that follows the disorder of the holes (in this respect, see the definition of the nonlinear Kerr coefficient in supplemental of \cite{Colman10}), we infer that the random process ruling the linear coupling coefficient also holds for the nonlinear coefficient. Specifically, in our model, this is taken into account by setting $\Psi_{\gamma}(z) = (\gamma/\Gamma) \Psi(z)$ and then $\sigma_{\gamma} = (\gamma/ \Gamma) \sigma_{\Psi}$.  
\begin{figure}[t!] 
{\includegraphics[width=8.5cm]{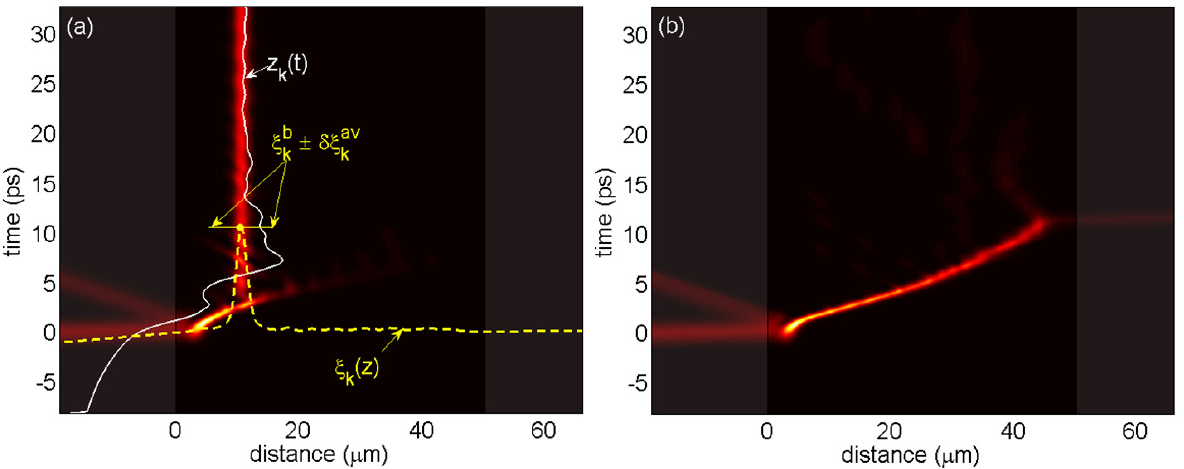}}
\caption{(Color online) NLCME simulations. (a) Example of application of metrics for GS in localized transport: soliton trajectory (white curve), soliton energy path (yellow dashed curve), and barycenter (yellow dot). (b) Example of GS in ballistic transport.} 
\label{soliton_trajectory}\end{figure}.
To describe wave propagation across the disordered lattice, we resort to an effective-particle approach \cite{Tsoy, deSterke, Kartashov_PRE_2005} conveniently modified for the soliton dynamics. In the specific, we define the weighted trajectories
\begin{equation}\label{eq3}
z_k(t)= \frac{\int z\left( |E_+|^2+v_r|E_-|^2\right)dz}{\int \left( |E_+|^2+v_r|E_-|^2\right)dz}
\end{equation}
and the soliton velocities
\begin{equation}\label{eq4}
v_k(t) = \frac{d z_k(t)}{dt}
\end{equation} 
Moreover, in order to track the wavepacket center of mass and intercept the localization point along the waveguide, we evaluate the soliton energy path as follows
\begin{equation}\label{eq5}
\xi_k(z) =  \frac{z\int \left( |E_+|^2+v_r|E_-|^2\right)dt}{ \left[\int \left( |E_+|^2+v_r|E_-|^2\right)dt \right]_{max}}
\end{equation}
that allows us to define the soliton barycenter:
\begin{equation}\label{eq6}
\xi_k^b= [\xi_k(z)]_{max}
\end{equation}
and the soliton path displacement:
\begin{equation}\label{eq7}
\delta \xi_k^{av} = \frac{1}{4L} \int \left( \xi_k(z) - \xi_k^b\right)^2dz
\end{equation}
In Eqs. (\ref{eq3}) and (\ref{eq5}) the backward wave is suitably attenuated by the ratio $v_r = V_i/V_g$ as the main intensity transport is ascribed to the forward wave.
If the most part of the flow light escapes the waveguide, the trajectory in Eq. (\ref{eq3}) exceeds the waveguide length, then the soliton velocity is derived by the slope of the trajectory. On the contrary, when localization happens, the instant $t_b$ of localization occurrence is exactly captured by the soliton barycenter in Eq. (\ref{eq6}), allowing us to evaluate the time averaged velocities 
\begin{equation}\label{eq8}
v_k^{av}=\frac{1}{T_b}\int_{T_b} v_k(t)dt
\end{equation}
and the velocity displacements
\begin{equation}\label{eq9}
\delta v_k^{av} = \frac{1}{4T_b} \int_{T_b} \left( v_k(t) - v_k^{av}\right)^2 dt
\end{equation}
where $T_b$ indicates the temporal interval beyond the instant $t_b$.
Statistical averaging over $N$ realizations these quantities, provides characterization of soliton localization. In particular, the localization length along with its squared deviation are obtained by 
\begin{equation}\label{eq10}
l_{loc} = <\xi_k^b>=\frac{1}{N}\Sigma_k \xi_k^b \quad \delta l_{loc} =  <\delta \xi_k^{av}> = \frac{1}{N}\Sigma_k \delta \xi_k^{av}
\end{equation}
while the mean soliton velocity and its squared deviation are given by 
\begin{equation}\label{eq11}
v_{m} = <v_k^{av}>= \frac{1}{N} \Sigma_k v_k^{av} \quad \delta v_m =  <\delta v_k^{av} > = \frac{1}{N}\Sigma_k \delta v_k^{av} 
\end{equation}
These metrics result to be a simple and well suited way to characterize localization especially for pulses impressed below the waveguide cut-off, namely Gap-Solions (GS), where alternative definitions can not be applied. As an example, Fig. \ref{soliton_trajectory} illustrates the use of these metrics for a pulse excited at $\omega=0.8\omega_b$ ($0.8\%$ above the semigap width), and the power (velocity) dependence of GS localization. 
As analyzed in the following, soliton can experience a transition from localized to ballistic transport by increasing the intensity of the excitation. In fact, the first pulse, launched at relative low power ($P=136.4$ W), highlights the localized regime (Fig. \ref{soliton_trajectory}(a)), while the second pulse, injected at higher power ($P= 143.6$W), manifests a ballistic transport (Fig. \ref{soliton_trajectory}(b)).\\

\begin{figure}[t!] 
{\includegraphics[width=9cm]{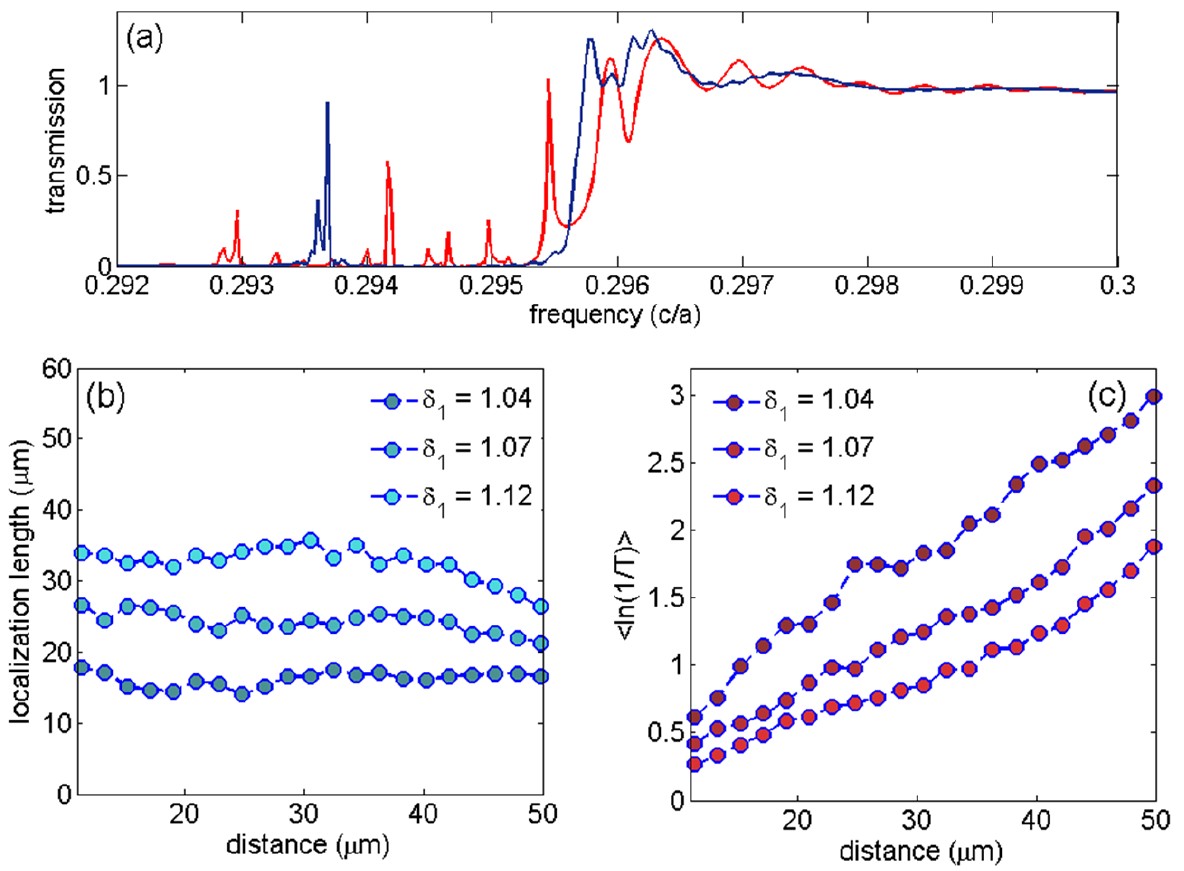}}
\caption{(Color online) Disordered PhC-WG analyzed by linear 2D FDTD simulations. (a) PhC-WG Transmission evaluated for two independent realization of disorder. (b)
Localization length $l_{loc}(z) = z/<\ln(1/T)>$ evaluated for different point along the waveguide. In (c) are shown the corresponding values of the ensemble average transmission $<\ln(1/T)>$.} 
\label{localization_fdtd}\end{figure}

\begin{figure}[t!] 
{\includegraphics[width=9cm]{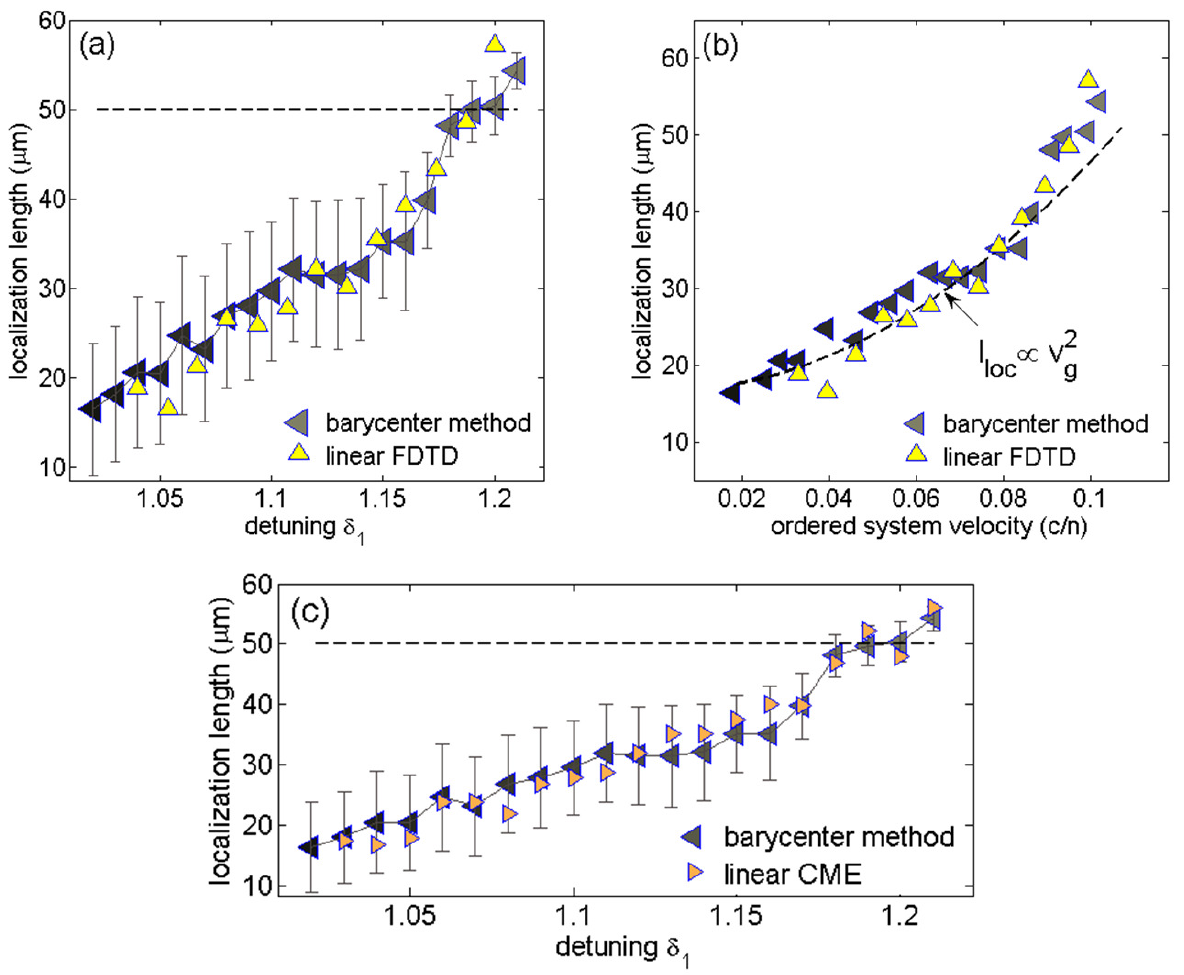}}
\caption{(Color online). Dispersive behavior of localization length in linear regime. (a)-(b) Localization length calculated by means of the barycenter method (gray triangles) and evaluated as $l_{loc}(z) = z/<\ln(1/T)>$  at $z=50\mu$m (yellow triangles) by FDTD method. In (c) we compare the localization length as $l_{loc}(z) = z/<\ln(1/T)>$ obtained by integrating in CW the linear coupled mode equations (pink triangles) and the outcomes of the barycenter method (gray triangles).} 
\label{localization_linear}\end{figure}

{\bf Validation of the barycenter method in linear regime}\\
The statistical quantity $\sigma_{\Psi}$ is obtained by inspection of the disordered structure in 2D-FDTD simulations. 
The FDTD approach, in fact, gives a reasonable description in 2D employing  effective index in the vertical direction and perfect matched layer (PML) absorbing boundary conditions in the plane of the PhC. In particular, a linear effective index $n_{eff}= 2.465$ (instead of the bulk value $n=3.17$) gives a cut-off frequency which matches the value $\omega_b$ from the 3D calculation.
Disorder is introduced by randomly and independently varying the holes radius around the mean nominal value $r=0.3a$ with a statistical Gaussian distribution of standard deviation $\sigma_r =0.0052a$ corresponding to a perturbation of $2.5$ nm.  Since a deviation of $2.5$ nm is smaller than the spatial resolution, we resort to a homogenization technique \cite{homog1, homog2} in the arrangement of the hole sides for the two electrical components of the TE-polarized light.
We simulate $N=50$ independent disordered realizations of the same waveguides. In particular, since the simulated waveguide has not terminations (cleaved or access-ridge facets), its spectrum is free from high-frequency oscillations corresponding to Fabry-Perot fringes, while sharp peaks appear at small frequency just below the cut-off frequency as a signature of Anderson localization \cite{Chabanov, Genack_Chabanov, Garcia, Mazoyer_Lalanne}. The sharp spikes due to disorder are evident in Fig. \ref{localization_fdtd}(a) in which we report the calculated PhC-WG transmission for two different instances of disorder.
Because of disorder, we also observe a slight shift of the cut-off. By evaluating the standard deviation of the frequency cut-off $\sigma_c =\sqrt{\frac{1}{N}\Sigma_k(f_k-f_c)^2}$, being $f_c$ the cut-off of the unperturbed PhC-WG and $f_k$ the cut-off of the $k$-th realization with disorder (estimated as the first zero of transmission below the pass-band), we find out a normalized (units of $c/a$) $\sigma_c = 2.82 \times 10^{-4}$. Finally, by looking for the $\Gamma$ distribution yielding the same cut-off standard deviation at the Brillouin edge $K=0.5$ we obtain $\sigma_{\Psi} = 0.14 \Gamma$.
.
We validate our barycenter method applied in linear regime of the system (\ref{eq1}) by comparison with 2D-FDTD linear simulations performed over a $50\,\mu$m long PhC-WG, for $50$ independent realizations of the disorder.   
By calculating the waveguide transmission $T$ evaluated as the DFT (Discrete Fourier Transform) for a given frequency and by collecting the ensemble average $<\ln(1/T)>$ at different distances $z$, we obtain the  localization length as defined in literature: $l_{loc}(z) = z/<\ln(1/T)>$ \cite{Lalanne_Melloni, Baron}.
Figs. \ref{localization_fdtd}(b) and (c) show the localization length and the averaged quantity $<\ln(1/T)>$ varying along the PhC-WG for different normalized detuning $\delta_1$ measured in the rest (lab) frame [for details see \cite{Malaguti}]. The barycenter method applied over $100$ instances of disorder in linear regime of the system (\ref{eq1}), gives the localization length as shown by gray triangles in Figs.\ref{localization_linear}. Specifically, in Fig. \ref{localization_linear}(a) we report the localization length as a function of the normalized detuning $\delta_1$. Here the black-dashed line stands for the waveguide termination at $50\,\mu$m, and all values located above this line must be intended as delocalized. More intuitively, localization should be given as a function of the group velocity (or equivalently group index) at a given detuning, provided that the group velocity is computed in the ordered structure where it keeps its meaning. This is shown in Fig. \ref{localization_linear}(b) where the localization length is reported versus group velocities (in units of $c/n$) corresponding to detunings in Fig. \ref{localization_linear}(a), but evaluated in the ideal ordered system.
In Figs. \ref{localization_linear}(a)-(b), by comparing the localization length $l_{loc}(z)$ at $z=50\,\mu$m obtained by FDTD simulations (yellow triangles) with the barycenter of the wave resulting from the coupled mode method (gray triangles), we show a good agreement between the two approaches.
.
To further confirm these results, we solve the system (\ref{eq1}) in linear regime for continuous waves ($\partial /\partial t =0$), and we calculate the localization length with the standard approach by averaging the logarithm of the transmission over $500$ realizations. Fig. \ref{localization_linear}(c) compares these results (pink triangles) with the outcomes of the barycenter method (gray triangles).
Noticeably, from this analysis the dispersive nature of the localization length is highlighted, implying the scaling law according to which the transition from localized to ballistic regime goes as the square of the group velocity \cite{Hughes_PRL_2005, Hughes_PRB_2009, Wang, Kuramochi}. This is established by the overlap of the values of localization length versus group velocity with the interpolated curve $l_{loc} \propto v_g^2$ as shown by the black-dashed curve in Fig. \ref{localization_linear}(b), where the proportionality turn out to be in good agreement with the approximated formula ($7$) in \cite{Wang}, in which a PhC-WG very similar to that used here is analyzed.\\

\begin{figure}[t!] 
{\includegraphics[width=8cm]{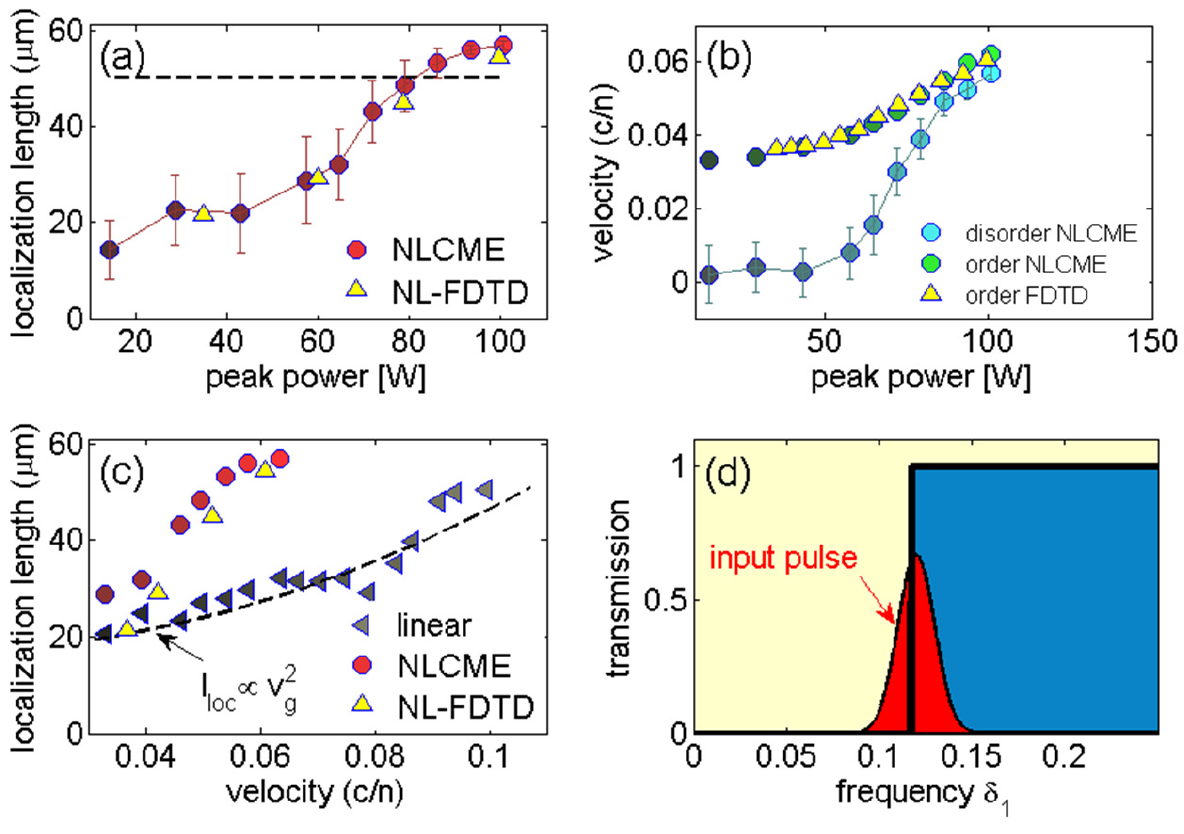}}
\caption{(Color online)}
\label{out_gap}\end{figure}.

{\bf Out-Gap dynamics}\\
In the following we address the soliton propagation close the cut-off frequency, where the impact of disorder pose major limitations in linear regime especially to achieve very slow light. In order to do that, we calculate the ensemble averaged quantities in Eqs. (\ref{eq10})- (\ref{eq11}) performing $100$ realizations of the disordered PhC-WG. In particular, we operate under normal experimental conditions by impressing a Gaussian input pulse with variable input power peak $P$, FWHM $2.2$ ps and detuned by $1.02\omega_b$ ($2\%$ above the semigap width), corresponding to a group index of almost $n_g=25$. Fig. \ref{out_gap}(d) sketched the PhC-WG transmission (blue) and the spectrum of the input pulse (red) for normalized frequency $\delta_1$ [for details see \cite{Malaguti}]. 
In Figs. \ref{out_gap}(a)-(b) we report, for increasing input peak power, the localization length (red circles) and the averaged velocity $v_m$ (blue circles), along with their standard deviation (error-bar) as obtained from Eq.(\ref{eq10}) and Eq.(\ref{eq11}), respectively. Relying on the NLCME equations, in Fig. \ref{out_gap}(b) we compare the soliton velocity in the ordered system (green circles) with the mean velocity in the disordered one (blue circles). We find that as the input power increases, the nonlinear waves become delocalized, while their averaged velocities approach the ideal ones. 
In order to test the validity of these results, we employ the time-domain planar 2D-FDTD method accounting for pure instantaneous Kerr effect. 
By following \cite{Malaguti} we integrate the Maxwell equations for the normalized fields in a medium with nonlinear effective index implicitly defined by the effective thickness of the slab.
We excite the PhC-WG with a field source centered on the waveguide axis and impressing a Gaussian profile in both $x$ direction and time, having a FWHM equal to that used in the reduced model. First of all, we analyze the ordered structure, and we obtain a reasonable agreement between the two methods. Indeed, as summarized in Fig. \ref{out_gap}(b), for the ordered waveguide the nonlinear FDTD outcomes (yellow triangles) result to be comparable with the NLCME ones (green circles). 
Secondly, we perform nonlinear FDTD simulations by launching the same pulse as above into the disordered PhC-WG. To take into account soliton localization, we evaluate the wavepacket center of mass by collecting for each time the $H_y$ component traveling toward the central axis $x_c$ of the waveguide then, in a way similar to that given by Eqs. (\ref{eq7})-(\ref{eq8}),  we calculate the soliton energy path as $\xi_k(z) = z[\int |H_y(x_c, z, t)|^2 dt]/[\int |H_y(x_c, z, t)|^2 dt]_{max}$ and the soliton barycenter as $\xi_k^b = [\xi_k(z)]_{max}$ for the k-th instance. Statistically averaging these quantities over $30$ independent realizations of disorder, we find out the localization values for increasing power. As shown in Fig. \ref{out_gap}(a) the FDTD results (yellow triangles) fit quite well the NLCME predictions. 
In order to demonstrate the ultimate improved robustness to disorder of solitons with respect to the linear regime, we compare the linear and nonlinear wave localization versus group velocities calculated in the ordered waveguide. In fact, we recall that despite in a disordered waveguide the group velocity of the Bloch mode loses its meaning near the waveguide cut-off, in an ideal waveguide it remains properly defined for both, the linear and nonlinear regime. Fig. \ref{out_gap}(c) shows that, while linear waves delocalization scales according to the $v_g^2$ law, for solitons the transition from localized to ballistic regime goes faster than the square of the group velocity, proving their enhanced robustness against disorder with respect to the linear regime. 
Importantly, we guess that, by virtue of the nonlinear modulation of the refractive index, the superiority of nonlinear waves can be ascribed to their capability of release themselves once trapped in spurious cavities created by disorder. From this standpoint, pulses at high optical intensity, relying on a sort of self-detuning, are more likely to escape these cavities than linear waves.
Close to the cut-off, this is obtained for relatively low input power ranging from $P \sim 10$ W to $ P \sim 100$ W. By further increasing the input power, we find that a multisolitonic regime takes place. On the contrary, at very low optical intensity, although GVD-dispersion dominates, the waveguide acts as a filter for a such (relatively) wide-band pulse and, by reshaping the wavepacket, allows the access through the waveguide only for the portion of the spectrum that lies above the cut-off, i.e. for frequency components that, once re-collected, show a group velocity that overcome the velocity of the excited pulse. This is the reason for which the nonlinear curve in Fig. \ref{out_gap}(c), even when GVD dominates, surpasses the linear one, being this latter evaluated in CW regime.\\

\begin{figure}[t!] 
{\includegraphics[width=9cm]{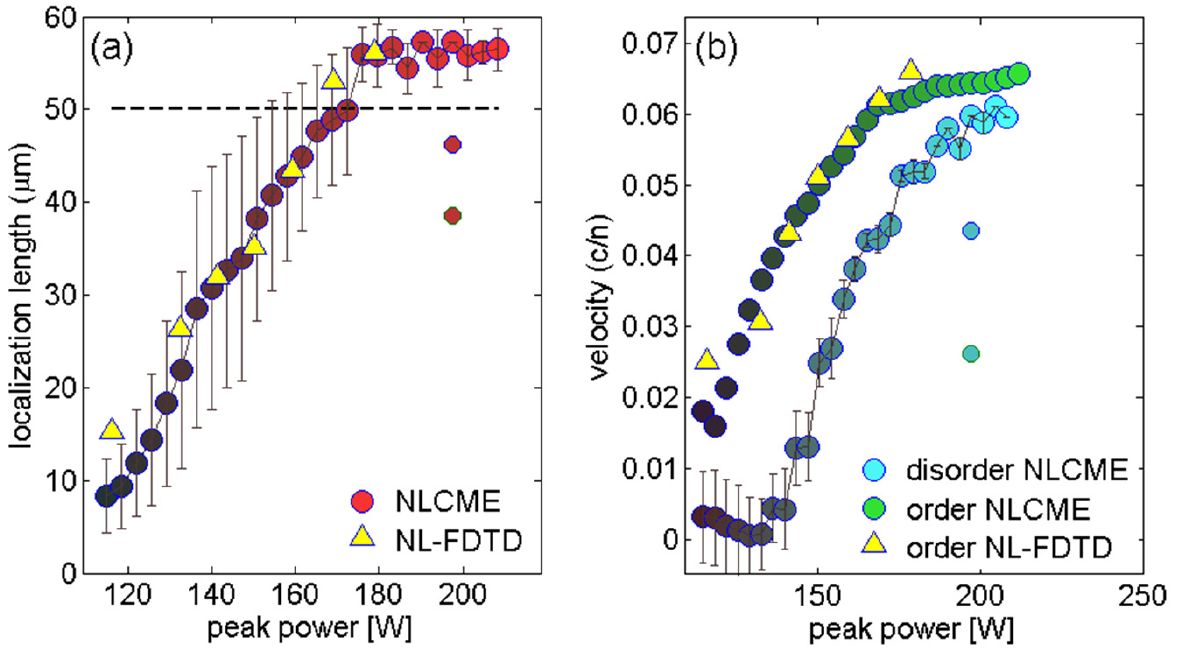}}
\caption{(Color online) Soliton delocalization for increasing input power. (a) Localiztion length versus input power given by NLCME (red circles) and FDTD (yellow triangles) simulations. (b) The mean velocity $v_m$ (blue circles) obtained by NLCME calculations with disorder approaches the ideal one evaluated in the ordered PhC-WG by means of NLCME (green circles) and FDTD (yellow triangles). Smaller circles are for increased disorder: $\sigma_{\Psi} = 0.18\, \Gamma,0.22\, \Gamma $ obtained in NLCME simulations.} 
\label{localization_det_0p8}\end{figure}

{\bf In-gap dynamics}\\
The capability to high control and tuning of group delay is much more effective where the linear transmission is forbidden, i.e. below the PhC-WG cut-off \cite{Malaguti}. However, for in-gap solitons (GS), the impact of disorder must be accurately assessed. Clearly, in the in-gap case, the only way through which this can be provided is by characterizing localization with the barycenter method, since the alternative standard calculation (localization obtained from $<\ln(1/T)>$) fails. 
The results are shown in Fig. \ref{localization_det_0p8} in which we still compare NLCME and FDTD outcomes, demonstrating transition from localized to ballistic regime for increasing power, although, in this case, larger peak powers in the range of hundred
Watts are required to excite the nonlinear modes 
.
Clearly, increased disorder much more affects the localized-to-ballistic transition. This is highlighted in Fig. \ref{localization_det_0p8} where small circles indicates instances for $\sigma_{\Psi} = 0.18\,\Gamma, 0.22 \,\Gamma$, i.e. standard deviations of $3.2$ nm and $3.6$ nm for the holes radius, respectively. 
\begin{figure}[t!]
{\includegraphics[width=8cm]{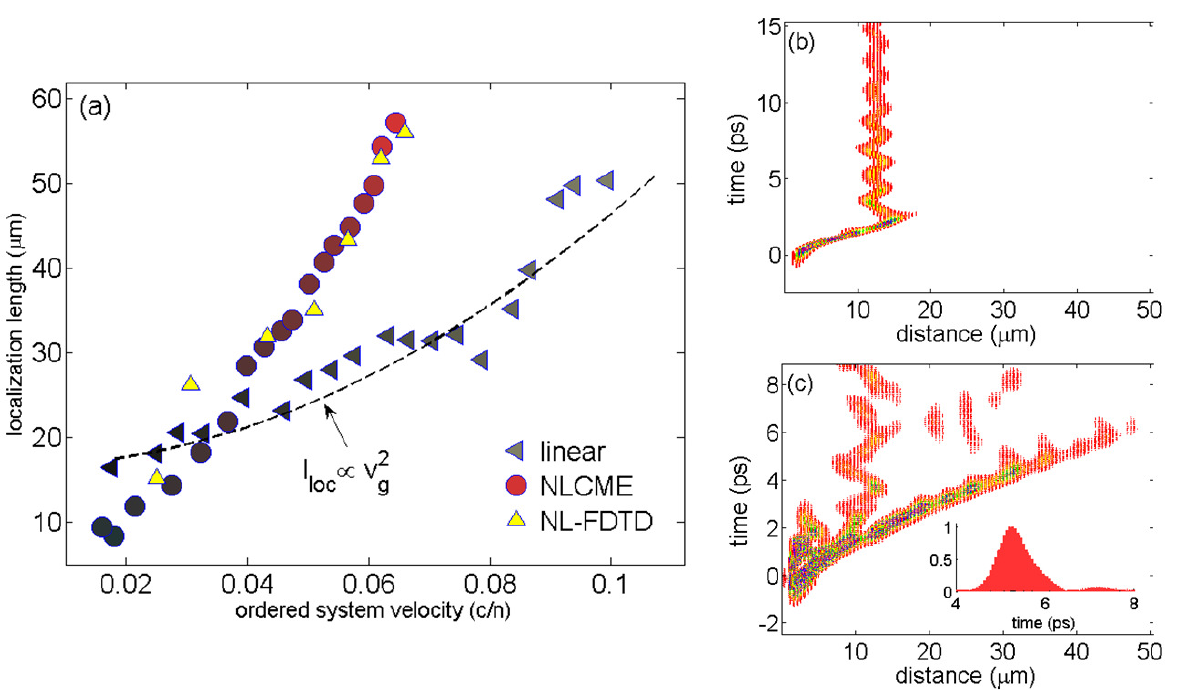}}
\caption{(Color online) (a) Comparison between the in-gap ($\delta_1 = 0.8$) and the linear localization dynamics as a function of the group velocity (units of c/n)evaluated in ordered systems. Transition from localized (b) to ballistic (c) regime as obtained in nonlinear FDTD simulations.}
\label{localization_compare}\end{figure}.
We point out that, as evident in Fig. \ref{localization_compare}(a), there is a threshold velocity at which the nonlinear localization curve overcomes the linear one. This is due to the fact that for not sufficiently high input power, the nonlinear wave is slowed down by dominant GVD-dispersion rather than accelerated by nonlinear group velocity enhancement. This latter is proved by nonlinear FDTD simulations in which the wavepacket broaden during the propagation (not reported in figure). 
Figs. \ref{localization_compare}(b)-(c) show the transition from localized (Fig. \ref{localization_compare}(b)) to ballistic (Fig. \ref{localization_compare}(c)) regime as obtained from FDTD simulations, for input power $P=132.5$ W and $P=178.7$ W, respectively. From FDTD results, unlike predicted by the NLCME model (see Fig. \ref{soliton_trajectory}), the pulse in ballistic regime loses part of its energy in some points of the waveguide. This is because of in a planar structure the disorder is not flattened in one dimension as in the reduced model. However, as shown in the inset of Fig. \ref{localization_compare}(c), the snap-shot of the pulse at distance $z=40\,\mu$m still have a Gaussian profile, that is a signature of ballistic transport \cite{Lalanne_Melloni}.

\end{document}